\documentclass[useAMS,usenatbib]{mn2e}
\usepackage{graphicx}
\usepackage{float}
\usepackage{hyperref}
\usepackage{color}
\usepackage{amsmath} 
\usepackage{amssymb} 

\newcommand{\aap}{A\&A}
\newcommand{\apj}{ApJ}
\newcommand{\apjs}{ApJS}
\newcommand{\apjl}{ApJL}

\newcommand{\mnras}{MNRAS}

\newcommand{\pasp}{PASP}
\newcommand{\aj}{AJ}

\title[Modeling the connection between ultraviolet and infrared galaxy populations across cosmic times]{Modeling the connection between ultraviolet and infrared galaxy populations across cosmic times}

\author[E.~Bernhard et al.]
{\parbox{\textwidth}{E.~Bernhard,$^{1,2}$\thanks{E-mail: \texttt{ebernhard1@sheffield.ac.uk}}
M.~B\'ethermin,$^{1,3}$
M.~Sargent,$^{1,4}$
V.~Buat,$^{5}$
J.R.~Mullaney,$^{2}$
M.~Pannella,$^{1}$
S.~Heinis,$^{5,6}$ and
E.~Daddi$^{1}$ }\vspace{0.4cm}\\\
\parbox{\textwidth}{$^{1}$Laboratoire AIM-Paris-Saclay, CEA/DSM/Irfu - CNRS - Universit\'e Paris Diderot, CEA-Saclay, pt courrier 131, F-91191 Gif-sur-Yvette, France\\\
$^{2}$Department of Physics $\&$ Astronomy, University of Sheffield, Sheffield S3 7RH, UK\\\
$^{3}$European Southern Observatory, Karl-Schwarzschild-Str. 2, 85748 Garching, Germany\\\
$^{4}$Astronomy Centre, Dept. of Physics \& Astronomy, University of Sussex, Brighton BN1 9QH, UK\\\
$^{5}$Aix Marseille Universit\'e, CNRS, LAM (Laboratoire d'Astrophysique de Marseille) UMR 7326, 13388, Marseille, France\\\
$^{6}$Department of Astronomy, University of Maryland, College Park, MD 20742-2421, USA}}

\begin{document}

\date{Accepted ??? Received ???; in original form ???}


\maketitle

\begin{abstract}
Using a phenomenological approach, we self-consistently model the redshift evolution of the ultraviolet (UV) and infrared (IR) luminosity functions across cosmic time, as well as a range of observed IR properties of UV-selected galaxy population. This model is an extension of the 2SFM (2 star-formation modes) formalism, which is based on the observed "main-sequence" of star-forming galaxies, i.e. a strong correlation between their stellar mass and their star formation rate (SFR), and a secondary population of starbursts with an excess of star formation. The balance between the UV light from young, massive stars and the dust-reprocessed IR emission is modeled following the empirical relation between the attenuation (IRX for IR excess hereafter) and the stellar mass, assuming a scatter of 0.4\,dex around this relation. We obtain a good overall agreement with the measurements of the IR luminosity function up to z$\sim$3 and the UV luminosity functions up to z$\sim $6, and show that a scatter on the IRX-M relation is mandatory to reproduce these observables. We also naturally reproduce the observed, flat relation between the mean IRX and the UV luminosity at L$_{\rm UV}>$10$^{9.5}$\, L$_\odot$. Finally, we perform predictions of the UV properties and detectability of IR-selected samples and the vice versa, and discuss the results in the context of the UV-rest-frame and sub-millimeter surveys of the next decade.
\end{abstract}

\begin{keywords}
galaxies: statistics -- galaxies: luminosity function, mass function -- ultraviolet: galaxies -- infrared: galaxies -- galaxies: evolution
\end{keywords}

\section{\label{sec:introduction} Introduction}

Galaxy evolution in the cosmological context is an important topic of modern astrophysics. In particular, processes that drive star formation and gas accretion within galaxies are not well understood. Observational constraints from deep surveys should provide insights for solving this complex problem. Obtaining accurate star formation rate (SFR) estimates for statistically significant samples of galaxies is thus particularly important. In deep photometric surveys, the rest-frame far-ultraviolet (hereafter UV) continuum emitted by blue, young and massive stars is a good probe of the SFR. Unfortunately, dust in star-forming regions absorbs a large fraction of the UV photons and reprocesses them into mid- and far-infrared (hereafter IR) emissions. The fraction of escaping UV photons can be estimated by computing the ratio between the IR and UV luminosity density ($\rm LD_{IR}/\rm LD_{UV} $), which are estimated integrating the UV \citep[e.g.][]{Arnouts2005,Cucciati2012} and IR \citep[e.g.][]{Le_Floch2005,Rodighiero2010,Magnelli2009,Casey2012,Gruppioni2013} luminosity functions. This quantity evolves from $\sim 60 \%$ in the local Universe up to $\sim 90 \%$ at $\rm z \sim 1.5$ \citep[e.g.][]{Takeuchi2005,Burgarella2013}.\\

To measure the SFR in galaxies, it is thus crucial to estimate accurately the fraction of UV light reprocessed by dust. The easiest method is to measure both unabsorbed UV and reprocessed IR emissions to trace the full intrinsic UV light emitted by young stars. Nevertheless, even the deepest and most recent IR surveys (e.g. \textit{Herschel}) are limited by the confusion phenomenon \citep{Dole2004,Nguyen2010} and the low sensitivity of the detectors as compared to the optical regime. Indeed, the majority of high redshift UV galaxies cannot be detected in IR. For these IR-undetected galaxies, the correction for dust attenuation has to be estimated using UV and optical data only. For instance, \cite{Meurer1999} proposed a relation, which links the attenuation to the UV continuum slope, calibrated using observations in the local Universe. This relation was tested up to $\rm z \sim 2$ \citep[e.g.][]{Buat2005,Reddy2010,Heinis2013a}. However, such analyses are either based on samples detected in both IR and UV, which could introduce a selection bias, or based on stacking analyses \citep{Reddy2010,Heinis2013a}, which does not allow to measure the scatter. Furthermore, the \cite{Meurer1999} relation may also be subject to aperture effects \citep{Overzier2011,Takeuchi2012}. The understanding of selection effects caused by the limited depth of IR observations and the very different dust attenuation observed in galaxies is thus crucial to estimate accurately the star formation rate density in the high redshift Universe.\\

These selection effects can be studied using phenomenological models. In particular, the 2SFM (2 star-formation modes, \citealt{Sargent2012,Bethermin2012c}) model was motivated by the finding of a strong correlation between the SFR and the stellar mass (M) of star-forming galaxies, called main sequence (MS), which evolves with redshift in normalisation \citep[e.g.][]{Noeske2007,Elbaz2007,Daddi2007}. This strong correlation suggests that the bulk of the stars are formed in galaxies following secular processes. \cite{Sargent2012} showed that the IR luminosity function can be reproduced well by adding to the MS a minor population ($\sim$3 \% of the whole population) of starburst (SB) that display a strong excess of SFR compared to the main sequence. This population is thought to have SFR driven by short-lasting, extreme events such as major mergers (\citealp{Elbaz2007,Tacconi2008,Elbaz2011,Daddi2010a, Genzel2010}). \cite{Bethermin2012c} show that galaxy number counts from the mid-infrared (MIR) to the radio can be reproduced considering distinct spectral energy distributions (SED) for MS and SB galaxies. Using a method based on abundance matching, \cite{Bethermin2013} have also shown that the 2SFM model efficiently predicts the large-scale fluctuations of the cosmic IR background produced by dusty, star forming galaxies, thereby probing the link between IR galaxies and dark matter halos. This modelling approach is thus very efficient for reproducing and interpreting IR observables. As \cite{Sargent2012} did not consider the unobscured component of SFR emitted in the UV by star forming galaxies, we propose in this paper to extend the model to the UV.\\

Our extension to UV emission requires a well-constrained prescription for how the intrinsic UV light emitted by young stars is redistributed between the unattenuated UV escaping from galaxies and the IR reprocessed by dust. Here we use the link between the stellar mass and the IR excess (IRX, defined as log(L$_{\rm IR}$/L$_{\rm UV}$), where L$_{\rm IR}$ and L$_{\rm UV}$ are respectively the IR and UV luminosities) found by several observational studies (\citealt{Pannella2009,Heinis2013b}; Pannella et al. in prep.). We note that we could also have used the relation between IRX and SFR. However, this relation is more difficult to measure in an unbiased and consistent way, because SFR estimates require, at least, UV flux measurements, and corrections for dust attenuation. The trends observed on samples detected in both UV and IR are less clear and have a large scatter \citep[e.g.][]{Buat2012}. In this paper, we show that an IRX-M relation based on the stacking analysis of \cite{Heinis2013b}, after addition of a dispersion, is capable of reproducing the statistical properties of UV galaxies.\\

In Sect.\,\ref{sect:moddescr} we describe the construction of our empirical model. In Sect.\,\ref{sect:lumfct}, we present our predictions for UV and IR luminosity function evolution and compare these with observations. In Sect.\,\ref{sect:irpropuv} we use our model to interpret key IR properties of UV-selected populations. In Sect.\,\ref{sect:pred} we present additional predictions of our model and discuss the UV-detectability of ALMA sources. We conclude in Sect\,\ref{sect:conc}.\\

In this paper, we assume a WMAP-7-year cosmology \citep{Larson2010} and a \cite{Salpeter1955} initial mass function (IMF), converting from a \cite{Chabrier2003} IMF where necessary.


\begin{table*}
\centering
\begin{tabular}{c c c}
\hline
\hline
Parameter & Description & Value $\pm$ Error \\
\hline
\multicolumn{3}{c} {Distribution of sSFR \citep{Sargent2012, Bethermin2012c}}\\
\hline
$\rm sSFR_{MS,0}$ & sSFR on the MS at z = 0 and M = $10^{11} \rm M_{\odot}$ (in $\log(\rm yr^{-1} )$) & $-10.2 \pm 0.1$\\
$\beta_{MS}$ & Slope of the $\rm sSFR-\rm M^{*}$ relation at a given redshift & $-0.2\pm 0.04$ \\
$\gamma_{MS}$ & Evolution of the normalisation of the MS with redshift & $3\pm 0.2$\\
$\sigma_{MS}$ & Width of the MS log-normal distribution (in dex) & $0.15 \pm 0.003$ \\
$\rm r_{SB_0}$ & Relative amplitude of SB log-normal distribution compared to MS & $0.012 \pm 0.003$\\
$\sigma_{SB}$ & Width of the SB log-normal distribution (in dex) & $0.20 \pm 0.07$ \\
$\rm B_{SB}$ & Boost of specific star formation rate in SB (in dex) & $0.6 \pm 0.07$\\
\hline
\multicolumn{3}{c} {Attenuation relation $\rm IRX=\alpha \log{M}+\rm IRX_{0}$ \citep{Heinis2013b}}\\
\hline
$\alpha$ & Slope of the $\rm IRX-\rm M^{*}$ relation & $0.71\pm 0.21$\\
$\rm IRX_{0}$ & Normalisation of the $\rm IRX-\rm M^{*}$ relation & $1.32 \pm 0.11$\,dex\\
\hline
\multicolumn{3}{c} {Mass function \citep{Ilbert2013}}\\
\hline
\multicolumn{3}{c} {All parameters and errors required to evaluate eq. \ref{eq:massdistrib} are listed in Table 2 of \cite{Ilbert2013}, converted to a Salpeter IMF}\\
\hline
\end{tabular}
\centering\caption{Central values and uncertainties on the parameters used in our model. These values come from previous works of \citealt{Bethermin2012c}, \citealt{Sargent2012}, \citealt{Heinis2013b}, and \citealt{Ilbert2013}.}
\label{tab:error}
\end{table*}

\section {Model description}

\label{sect:moddescr}

Our model is based on empirical relations derived from recent observational studies. We use the following steps to derive the evolution of the statistical UV and IR properties of galaxies with cosmic time:
\begin{itemize}
\item Our analysis is based on the observed mass functions of \cite{Ilbert2013} which allow us to fix the number of galaxies with a certain stellar mass in a given cosmological volume (Sect.\,\ref{sect:mf});
\item We then assume a SFR-M relation and a scatter around it following \citet{Sargent2012} to stochastically assign a SFR to each galaxy (Sect.\,\ref{sect:sfr});
\item Finally we apply the observed IRX-M relation of \cite{Heinis2013b} to predict the relative fraction of UV light that escapes star-forming galaxies and that is reprocessed by dust, respectively (Sect.\,\ref{sect:irxm});
\item Additional refinements to the model, e.g. the special treatment of the dust-attenuation in starbursts and the evolution of the IRX-M relation at z$<$1, are described in Sect.\,\ref{subsec:refinements}.
\end{itemize}

\subsection{Mass function}

\label{sect:mf}

As our starting point we use the mass functions of star-forming galaxies at different redshifts (0$<$z$<$4) to predict their expected number per comoving volume in various mass and redshift bins. Our analysis thus neglects the passive galaxy population. The version of the 2SFM model used by previous analyses was based on the evolving mass functions of \cite{Ilbert2010} which cover the redshift range of  $z\lesssim2$. These have now been extended to $z\sim4$ by \cite{Ilbert2013}. As such we use the full range of mass functions of this latter study, i.e., $0.2<z<4$. As well as extending to higher redshifts, \cite{Ilbert2013} also found steeper low mass tails at all redshifts, implying greater numbers of low mass galaxies per comoving volume. The parametric form of this mass function, as presented in \cite{Ilbert2013} and implemented in this model version, is given by:\\

\begin{equation}
	\begin{split}
	\phi(\rm M) ~\rm dlog(M)=\exp \left (-\frac{\rm M}{\rm M^{*}} \right )~ \left [\phi^{*}_1\left (\frac{\rm M}{\rm M^{*}} \right)^{\alpha_1}+\phi^{*}_2 \left (\frac{\rm M}{\rm M^{*}} \right)^{\alpha_2} \right ]\\
	\times \left (\frac{\rm M}{\rm M^{*}} \right )\log(10) ~\rm dlog(M),
	\end{split}
\label{eq:massdistrib}
\end{equation}

\noindent where the values for $M^{*}$ (i.e., the location of the knee of the stellar mass function), $\phi^{*}_1$, $\phi^{*}_2$, $\alpha_1$ and $\alpha_2$ are given in Table 2 of \cite{Ilbert2013} for each redshift slice. Above $z=1.5$, $\phi^{*}_2$ is fixed to 0, thus, equation \ref{eq:massdistrib} becomes a simple Schechter function with parameters $\phi^{*}$ and $\alpha$. For redshifts lower than 0.2, the evolution of the \cite{Ilbert2013} mass-function parameters is extrapolated from the higher-redshift values. We note that our extrapolation is consistent with \cite{Baldry2012} at $z=0.06$ (i.e., the highest redshift of that study). Finally, we note that the redshift bins of \cite{Ilbert2013} are typically too broad for use in our model, leading to discontinuities in the galaxy redshift distributions and over- or under-estimates of source counts at the extremes of each redshift bins. To avoid this, we linearly interpolate their bins onto a finer redshift grid with logarithmic spacing $\Delta{\rm log}(z)$\,=\,0.05. The evolution of the mass function of star-forming galaxies used in our analysis is shown in Fig.\,\ref{fig:moddescr} (upper panel).


\subsection{Link between stellar mass and star formation rate}

\label{sect:sfr}

\begin{figure}
\includegraphics[width=8.5cm]{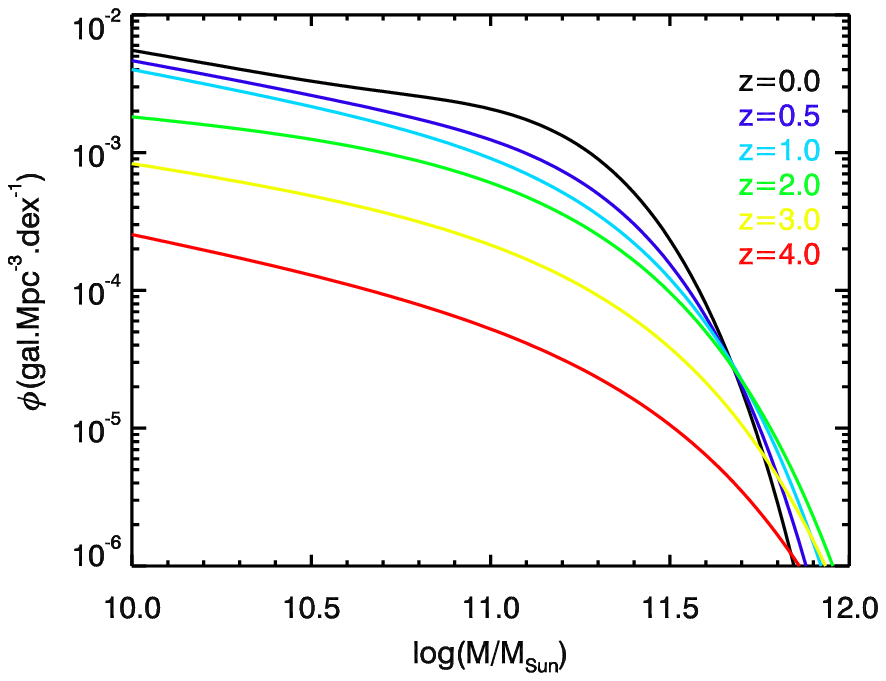}
\includegraphics[width=8.5cm]{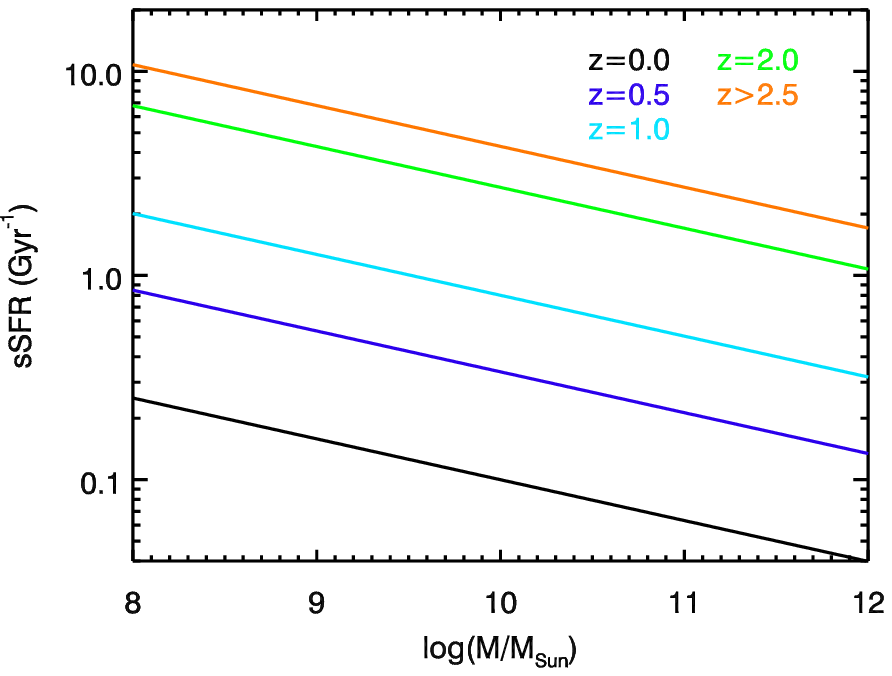}
\includegraphics[width=8.5cm]{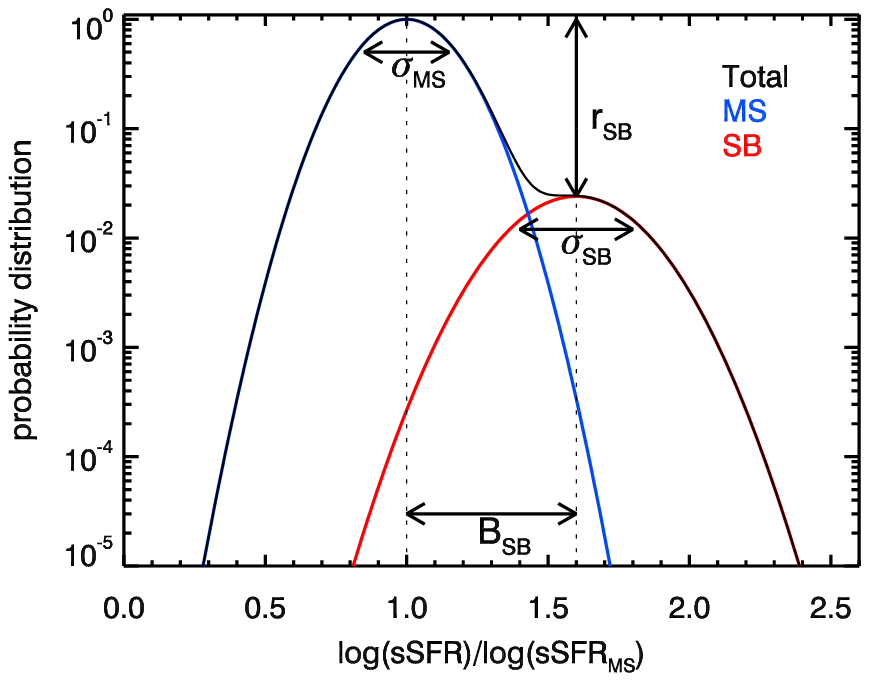} 
\caption{\label{fig:moddescr} \textbf{Upper panel:} Evolution of the stellar mass function of star-forming galaxies used in this paper. \textbf{Central panel:} Evolution of the mean sSFR-M relation (i.e. the main-sequence) as a function of redshift. \textbf{Lower panel:} Figure illustrating how the scatter around the main-sequence is parametrised in our model. $\rm r_{\rm SB}$ corresponds to the ratio between the amplitude of the peak of the SB and MS distributions.} 
\end{figure}

To derive the specific SFR (sSFR=SFR/M) for the galaxy population described by our chosen mass functions (see above) we use the same parametric representation of the main-sequence as \citet{Bethermin2012c}:

\begin{equation}
	\begin{split}
	\rm sSFR_{\rm MS}(\rm M,\rm z) = \rm sSFR_{\rm MS,0} \left(\frac{\rm M}{10^{11}\rm M_{\odot}}\right)^{\beta_{\rm MS}}\left(1+\rm min(\rm z,\rm z_{\rm evo})\right)^{\gamma_{\rm MS}},
	\end{split}
	\label{eq:ssfrms}
\end{equation}

\noindent where $\textrm{sSFR}_{\rm MS}$ is the sSFR of a $10^{11} \rm M_{\odot}$  galaxy lying exactly at the center of the main sequence, log$(\textrm{sSFR}_{\textrm{MS},0}) = -10.2$\,yr$^{-1}$, $\beta_{\rm MS}=-0.2$, $z_{\rm evo} = 2.5$, and $\gamma_{\rm MS}$ = 3. These values are based on observational calibrations as discussed in \citet{Sargent2012} and \citet{Bethermin2012c}. The chosen scenario for the SFR-M relation, described by Eq.\,2 (i.e. a rising power law, following by a plateau beyond z=2.5), is illustrated in Fig\,\ref{fig:moddescr} (central panel). Recent studies (e.g. \citealt{De_Barros2014}) suggest that, once the impact of nebular emission lines on the near-IR photometry (which leads to an overestimate of the stellar mass) is correctly taken into account, sSFRs continue to increase beyond z=2.5. However, the mass functions of \citet{Ilbert2013}, the observational calibration of the sSFR evolution, and the measurements of the IRX-M relation have all been computed neglecting this phenomenon. Since SFRs and IRXs are oberved properties, updating the masses has a proportionnal impact on these relationships, which cancels when we assume these updated masses in our analysis. As such, correcting for nebular emission has no impact on the derived UV and IR galaxy properties.\\

To describe the dispersion around this relation, we used the probability distribution of \cite{Sargent2012} which incorporates both MS and SB galaxies and is based on the empirical results of \cite{Rodighiero2011}. The result of \cite{Sargent2012} was based on a sample of $z\sim2$ star forming galaxies with $M>10^{10}M_{\odot}$, and is parametrized as a double log-normal law decomposed into MS and SB components:\\

\begin{equation}
	\begin{split}
	\rm p(\log(\rm sSFR))\propto \exp({-\frac{(\log(\rm sSFR)-\log{\rm (sSFR_{MS})})^2}{2\sigma_{MS}^2}})\\
	+\rm r_{SB}\exp({-\frac{(\log(\rm sSFR)-\log{\rm (sSFR_{MS})}-\rm B_{SB})^2}{2\sigma_{SB}^2}})
	\end{split}
	\label{eq:lognormal}
\end{equation}

\noindent where $\sigma_{MS}$ and $\sigma_{SB}$ are the standard deviations of MS and SB log-normal distributions respectively, ${\rm sSFR_{MS}}$ is the central sSFR value for MS galaxies at $z\sim 2$, $r_{SB}$ is the relative amplitude between SB and MS galaxies and $B_{SB}$ (interpreted as the average sSFR "boost" for SB galaxies in \citealt{Sargent2012}) is the ratio between the sSFR loci of SB and MS galaxies. This is illustrated by Fig.\,\ref{fig:moddescr} (lower panel).\\ 

As our model incorporates galaxies spanning a wide range of masses and redshifts we must modify the probability distribution of \cite{Sargent2012} to take this into account.  For this, we adopt the same assumptions as adopted by \cite{Bethermin2012c} to reproduce IR number counts. To summarise: 
\begin{itemize}
\item $\sigma_{\rm MS} = 0.15$~dex, $\sigma_{\rm SB} = 0.2$~dex, and $\rm B_{\rm SB}=0.6$ are kept constant throughout\footnote{Following \citet{Bethermin2012c}, the measured values of $\sigma_{\rm MS}$ and $\sigma_{\rm SB}$ are corrected for artificial broadening effects, already noted in \cite{Rodighiero2011}, namely (a) cosmological evolution within finite redshift bins, and (b) instrumental noise. $\sigma_{\rm SB}$ is slightly larger than $\sigma_{\rm MS}$ because of the scatter about the mean sSFR-boost observed for SB (see discussion in \citealt{Sargent2013})} (i.e., unchanging with respect to either mass or redshift) for simplicity. Furthermore, \citet{Sargent2012} found no evidence for evolution of these parameters;
\item ${\rm sSFR_{\rm MS}}$ evolves with redshift and mass as described previously (eq \ref{eq:ssfrms});
\item Following \cite{Bethermin2012c}, and consistent with the indications for somewhat lower starburst contribution at z=0 found in \cite{Sargent2012}, we set $\rm r_{\rm SB}\propto(1+\rm z)$ at $\rm z<1$, followed by a constant, $\rm r_{\rm SB}=0.024$ (\citealt{hopkins2010} and \citealt{Sargent2012}) at higher redshifts (Eq.\,3 in \citealt{Bethermin2012c}).
\end{itemize}

\subsection{\label{sect:IRX}Relation between stellar mass and dust attenuation}

\label{sect:irxm}

Dust-enshrouded star-forming regions absorb UV light emitted by young stars and emit it as IR radiation. As such, the relation between dust attenuation and stellar mass, hereafter called `attenuation relation', can be used to compute the ratio between the IR and UV luminosity for our star-forming galaxies, as shown in several studies (e.g. \citealt{Pannella2009}, \citealt{Heinis2013b}, and Pannella et al. in prep.). We used the following parametric form for the IRX-M relation:
\begin{equation}
\textrm{IRX}=\log(\rm L_{IR}/\rm L_{UV})=\alpha \left ( \log(\frac{\rm M}{ M_{\odot}}) - 10.35 \right ) +{\rm IRX}_{0},
\label{eq:irx}
\end{equation}
where $\alpha=0.71\pm 0.21$ and $\rm IRX_0=1.32\pm 0.11$ following \cite{Heinis2013b}, who do not see any evolution of this relation between z=1.5 and z=4. This relation was measured only for galaxies with M$>10^{9.5}\, \rm M_\odot$ while we here extrapolate it also to lower masses and beyond this redshift range. \cite{Heinis2013b} base their work on UV-detected populations but their findings are similar to new results obtained by Pannella et al. (in prep.) with a purely mass-selected sample of star forming galaxies.\\

From the SFR, we then derive the UV and IR luminosity emitted by the galaxy, assuming an energy balance between the SFR probed by unattenuated UV light and reprocessed IR, i.e. $\rm SFR=\rm SFR_{IR}+\rm SFR_{UV}$. The link between SFR and, IR or UV, luminosities is computed following \citet{Kennicutt1998}: $\rm SFR_{\rm IR}=\rm K_{IR} \times \rm L_{IR}$ with $\rm K_{IR} =1.7 \times 10^{-10}$\,M$_\odot$yr$^{-1}$L$_\odot^{-1}$ and $\rm SFR_{UV}=\rm K_{UV} \times \rm L_{UV}$ with $\rm K_{UV} = 2.8 \times 10^{-10}$\,M$_\odot$yr$^{-1}$L$_\odot^{-1}$. The IR luminosity is the total bolometric emission of the galaxy from 8\,$\mu$m to 1000\,$\mu$m rest-frame and the UV luminosity is the monochromatic luminosity at 150\,nm rest-frame. Combining these three equations and the IRX-M relation, we obtain:
\begin{equation}
	\rm L_{IR}=\frac{\rm SFR}{\rm K_{UV}}\frac{10^{IRX}}{1+\frac{\rm K_{IR}}{\rm K_{UV}}10^{IRX}} \, \,\textrm{and} \, \,  \rm L_{UV} = \frac{\rm SFR}{\rm K_{UV}}\frac{1}{1+\frac{\rm K_{IR}}{\rm K_{UV}}10^{IRX}}.
	\label{eq:lir}
\end{equation}


\subsection{Refinements}

\label{subsec:refinements}

The attenuation relation used in our model \citep{Heinis2013b} is derived from a stacking analysis and hence provides no information on the dispersion around this mean relation. This kind of measure is difficult to perform and could be strongly biased by selection effects. \citet{Buat2012} studied the IR emission of a sample of UV-selected galaxies at $z\sim1.5$ and found a mean relation between IRX and M that is fully consistent with the one of \citet{Heinis2013b} which we adopt for our predictions. \citet{Buat2012} measure the observed scatter around the mean relation to be 0.30 dex. This value must be considered a lower limit in view of the incompleteness of their sample, in which galaxies have to be detected both in the UV and IR. A value of  0.3 dex is also representative of the dispersion found on IRX at redshifts from 0 to 2 for a given UV luminosity (\citet{Heinis2013b} and references therein). Pannella et al. (in prep.) estimated a dispersion of 0.4 dex using the \textit{Herschel} detections up to z= 1.3 in the deepest regions of the H-GOODS survey. \cite{Wuyts2012} report that the dispersion of the attenuation relation is independent of stellar mass. We will thus assume a constant log-normal scatter of 0.4 dex (1 magnitude) versus M for simplicity.\\

\citet{Takeuchi2005} have found that the ratio between the IR and the UV luminosity density increases by a factor of $\sim$3 ($\sim$0.5\,dex) from redshift 0 to 1. The evolution of this integrated property cannot be reproduced by the simplest version of our model because of the \textbf{adopted} non-evolution of the IRX-M relation and the weak evolution of the mass function of star-forming galaxies \citep{Ilbert2013,Muzzin2013} below z=1. To solve this problem, we thus assume a slightly evolving normalisation of the IRX-M relation: 
\begin{equation}
	\rm IRX_0'=\rm IRX_0-0.5 \times (1-z) \hspace{1cm} \textrm{at} \, \, \, z<1.
	\label{eq:lowzcorrection}
\end{equation}
We could have also assumed an evolution of the slope of the relation, but we chose to use an evolving normalisation for simplicity. Indeed, \citet{Buat2013b} found lower attenuations at fixed mass in the local Universe. In a similar way, Pannella et al. in prep. found also a lower normalisation of the IRX-M relation in their z=0.5-1 bin than at z$>$1. This evolution of IRX at fixed mass can be also connected with the evolution of the average ratio between dust and stellar mass (B\'ethermin et al. in prep., Tan et al. in prep.), and the increase of disk-opacity observed by \citet{Sargent2010}, below z=1. This modification of the individual properties of galaxies correctly reproduced that evolution of LD$_{\rm IR}$/LD$_{\rm UV}$ at low redshift. We explicitly show how much this modification impacts the predictions of the model in Sect.\,\ref{sect:lfuv}. While \citet{Heinis2013b} find the attenuation relation to remain constant at high redshift, \citet{Cucciati2012} and \citet{Burgarella2013} report a decreasing integrated LD$_{\rm IR}$/LD$_{\rm UV}$ at z$>$2. These two observations do not contradict each other as the higher mean transparency of the Universe at high redshift is naturally recovered by our model because of the decrease of M$^\star$ at high redshift \citep{Ilbert2013,Muzzin2013}. The star formation budget at very high redshift is thus dominated by lower mass galaxies and the average attenuation is accordingly lower.\\

In the simplest implementation of our model SB galaxies are much less attenuated than MS galaxies with a similar SFR because of their lower mass. This disagrees with the observational results of e.g. \cite{Wuyts2011}. We hence adopt another prescription whereby starburst activity is taken to be fully dust-obscured. Specifically, we assume that star formation in SB galaxies is split between two processes such that SFR=$\rm SFR_{MS}+\rm SFR_{SB}$: extended star formation -- subject to the same mass-dependent attenuation (see eq. \ref{eq:irx}) as normal main-sequence galaxies -- and one or several compact SB regions, probably induced by major mergers, which are heavily dust-enshrouded and consequently emit only in the IR. We showed in \cite{Sargent2013} that it is possible to statistically establish by how much the activity of starbursting galaxies of a given sSFR-excess has been boosted with respect to their pre-burst state on the star-forming main sequence. For the present analysis, we use these sSFR-dependent boost-distributions \citep[see][Sect. 4.2.3]{Sargent2013} to compute the relative importance of the two star-formation modes for a given SB galaxy. The surplus SFR activity, $\rm SFR_{SB}$ (i.e. the difference between the SFR in the starburst state and the SFR prior to the onset of the burst phase), is assumed to be fully dust-obscured, while the remainder, $\rm SFR_{MS}$, is identified with the only partially obscured, extended component of the star-formation activity in the SB galaxy.


\begin{figure*}
	\centering
	\includegraphics[scale=0.95]{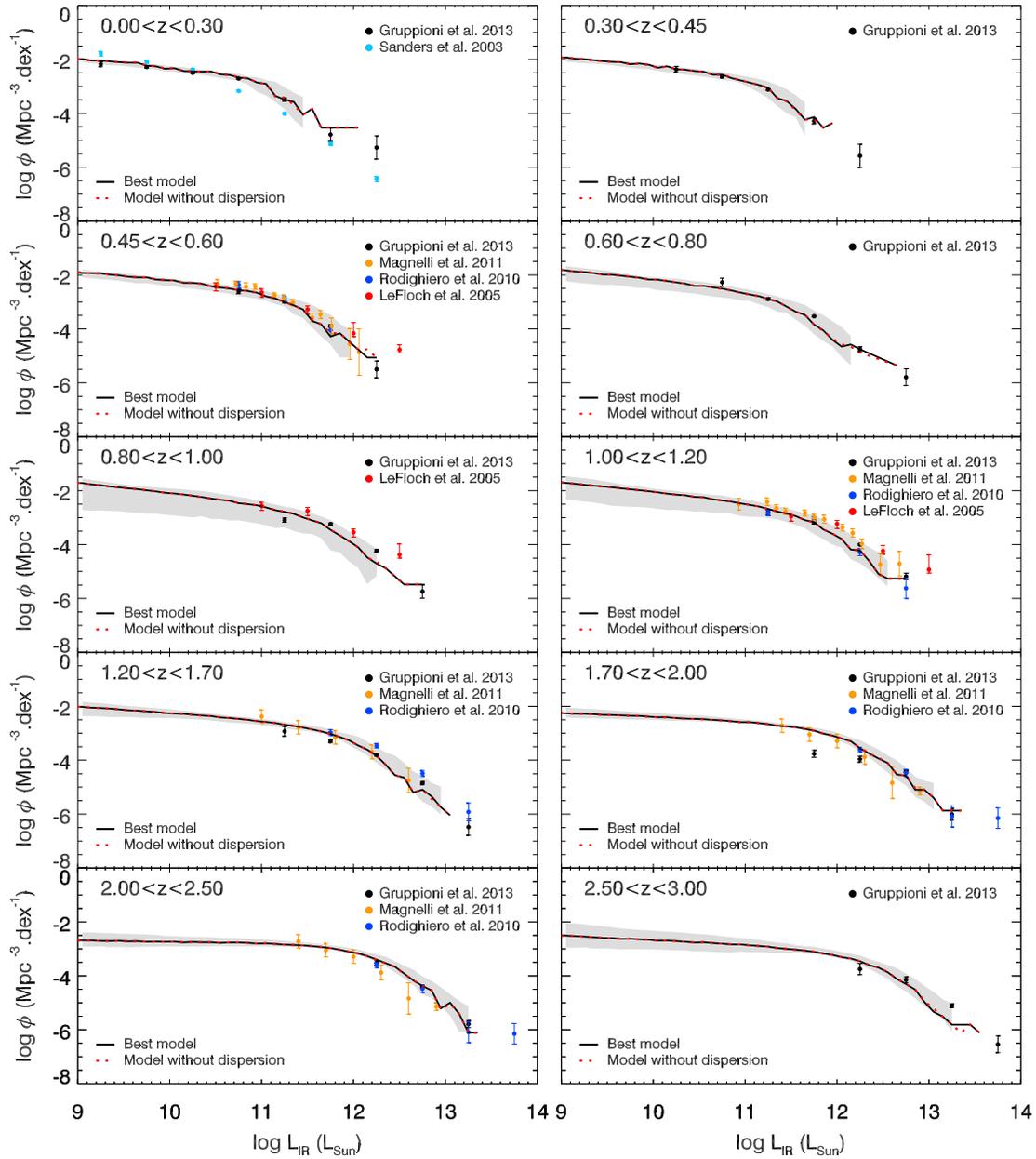}
	\caption{\label{fig:irlf} IR luminosity function from z$\sim$0 to z $\sim$3. Black solid lines represent the IR luminosity functions derived from our best model. Red dotted lines represents the same model, but without scatter on the attenuation relation. Grey areas represent the one sigma confidence region of our best model. Colored points are measurements from various observational studies \citep{Gruppioni2013, Magnelli2011, Rodighiero2010, Le_Floch2005, Sanders2003}.}
\end{figure*}

\begin{figure*}
	\centering
	\includegraphics[scale=0.95]{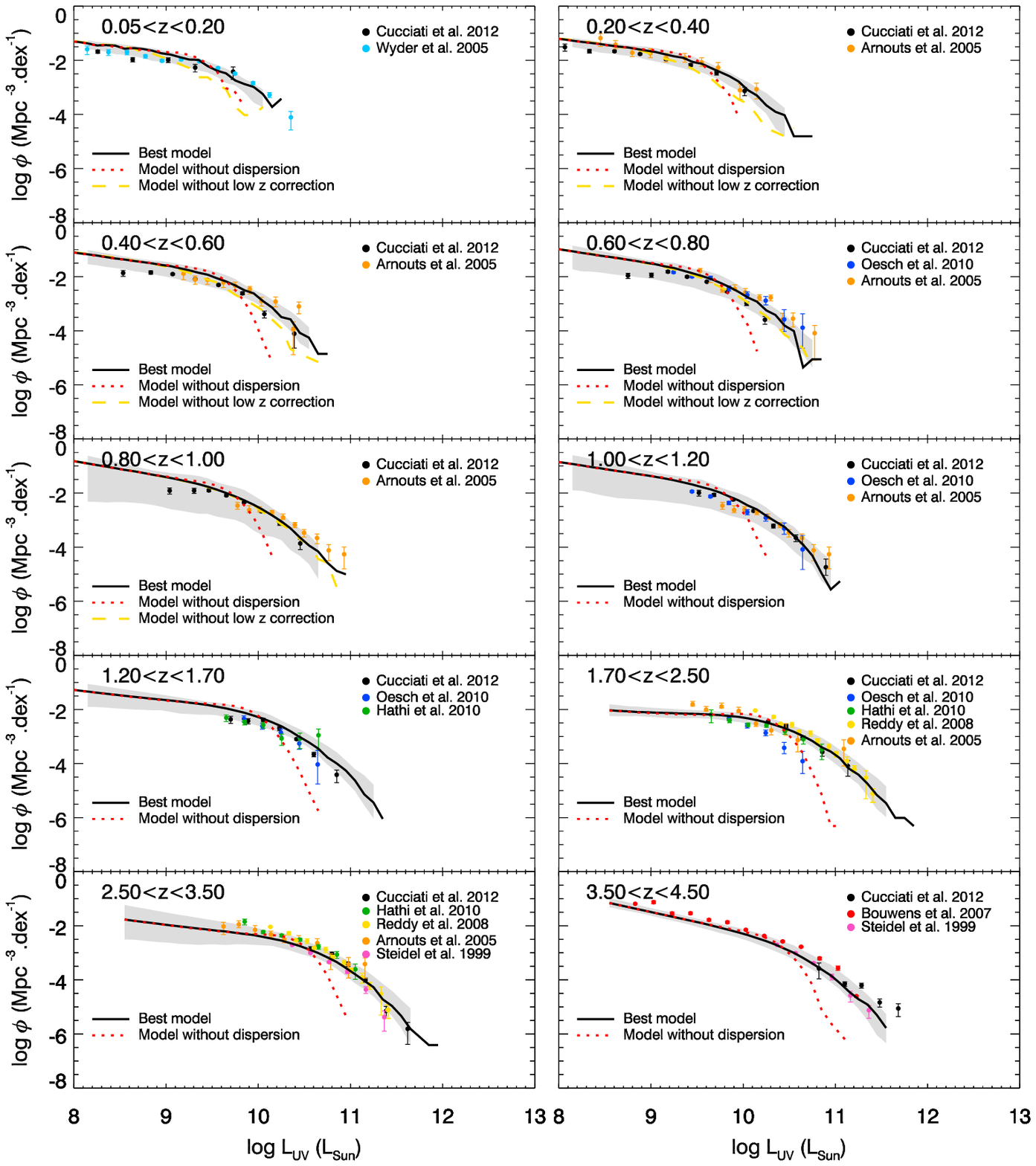}
	\caption{\label{fig:uvlf} UV luminosity function from z$\sim$0 to z $\sim$4. Black solid lines represent the UV luminosity function derived from our best model. Red dotted lines represent the same model, but without scatter on the attenuation relation. Yellow dashed lines represents the best model, but without evolution of the IRX-M relation at low redshift. Grey areas represent the one sigma confidence region of our best model. Colored points are measurements from various observational studies \citep{Cucciati2012, Hathi2010, Oesch2010, Reddy2008, Bouwens2007, Arnouts2005, Wyder2005, Steidel1999}}
\end{figure*}

\section{UV and IR luminosity functions}

\label{sect:lumfct}


\subsection{Computation of the luminosity function and uncertainties}

The computation of the IR and UV luminosity functions is difficult to perform analytically, because our model uses several scaling laws including a dispersion around them. We thus use the simpler method of constructing mock catalogues that we generate following the prescriptions detailed in Sect.\,\ref{sect:moddescr}. Luminosity functions are then derived by counting the number of galaxies in each redshift slice and luminosity bin, and dividing it by the volume of the redshift slice and the size (in dex) of the luminosity bin.\\

The mock catalogues are generated as follows:
\begin{itemize}
\item We generate a mock catalogue containing only M and z based on the mass function described in Sect.\,\ref{sect:mf} (Eq.\,\ref{eq:massdistrib}). In practice, we use a 2\,deg$^2$ field, to reach a compromise between sufficient statistics and computation time, and split it by redshift slices of $\Delta \log (1+z) = 0.05$. Mock galaxies are then generated based on the mass function at the center of each slice. Redshifts are drawn in the slice assuming a flat distribution. The number of objects predicted by the mass function diverges at low mass. Consequently, we apply a mass cut of 10$^6$\,M$_\odot$. We have checked that this low mass cut has no impact on the luminosity functions in the regime where data points are available. 

\item We assign a (s)SFR for each mock galaxy based on the prescriptions of the 2SFM model described in Sect.\,\ref{sect:sfr}. To do this, we first compute $\rm sSFR_{MS}$ from the M and z of each object using Eq.\,\ref{eq:ssfrms}. We then compute the SB fraction ($\rm f_{\rm SB}(\rm z)$) given by the ratio between the integration of the SB probability distribution (i.e. $2^{nd}$ part of Eq. \,\ref{eq:lognormal}) and the sum of the whole probability distribution (Eq.\,\ref{eq:lognormal}). Following $\rm f_{\rm SB}(\rm z)$, we can now draw a type (i.e. MS or SB). Then, we draw a sSFR for MS type along a gaussian centered on $\rm sSFR_{\rm MS}$ and with a $\sigma_{\rm MS}$ standard deviation; and for SB type along an other gaussian centered on $\rm sSFR_{\rm SB}=\rm sSFR_{\rm MS}+\rm B_{\rm SB}$ and with a $\sigma_{\rm SB}$ standard deviation.

\item We decompose the SFR in an UV and IR component. For each MS galaxy, we compute the mean expected IRX from M with Eq.\,\ref{eq:irx} and draw its actual IRX assuming a scatter of 0.4\,dex (1\,mag). We then compute the $\rm L_{UV}$ and $\rm L_{IR}$ using Eq.\,\ref{eq:lir}. For each SB object, we decompose the SFR into a MS and a SB component based on the boost-function formalism of \citet{Sargent2013}. The MS component is treated as previously explained. The IR luminosity emited by the SB component is assumed to be fully obscured, such that $\rm L_{IR,SB} = \rm SFR_{SB}/ \rm K_{IR}$.\\
\end{itemize}

We generated several mock catalogues. Our best model follows all the prescriptions listed in Sect.\,\ref{sect:moddescr}, including the refinements described in Sect.\,\ref{subsec:refinements}. We also generated a mock catalogue without scatter on the IRX-M relation and another one without low redshift modification of this relation. To derive confidence region for the best model, we generated 50 Monte Carlo mock catalogues using parameter values drawn randomly in their error region as summarised in Table\,\ref{tab:error}.\\


\subsection{Comparison with observed infrared luminosity functions up to z$\sim$3}

Using the same approach but neglecting attenuation, \cite{Sargent2012} were able to recover IR luminosity function evolution up to z$\sim$2. In this section, we compare the IR luminosity function evolution we obtain when including the attenuation with measurements at z$\sim$2 and beyond \citep[][, see Fig.\,\ref{fig:irlf}]{Sanders2003,Le_Floch2005,Rodighiero2010,Magnelli2011,Gruppioni2013}. We use the same redshift bins as \citet{Gruppioni2013}, whose measurements cover the largest redshift range. The confidence region of our model (grey area) agrees well with the data, except a mismatch with the measurements of \citet{Sanders2003}. These data are based on a very local sample and there is a significant evolution of the luminosity function up to z=0.3 (essentially a luminosity-evolution proportional to $(1+z)^3$). This tension disappears if we compute the luminosity function only up to z=0.05. We also tested the impact of the scatter in the IRX-M relation, which was not implemented in the \citet{Bethermin2012c} source counts model, and found this to be totally negligible (red dotted line). This is not surprising. Most of IR-detected galaxies are massive and their star-formation activity predominantly manifests itself as IR emission. Unattenuated UV light thus represents a modest fraction of the energy budget. Consequently, increasing or decreasing it by 0.4\,dex and removing or adding it to IR emission has thus a very modest impact. This also implies that it is irrelevant for the IR luminosity function whether we assume an evolving IRX-M relation at $z<1$ or not. In Fig.\,\ref{fig:irlf} we only plot luminosity functions for our realisations including this evolution.\\


\subsection{Comparison with observed UV luminosity functions at $z<4$}

\label{sect:lfuv}

In Fig.\,\ref{fig:uvlf}, we compare our model predictions with the UV luminosity functions measured by various authors \citep{Cucciati2012, Hathi2010, Oesch2010, Reddy2008, Bouwens2007, Arnouts2005, Wyder2005, Steidel1999}. We used the same redshift bins as \citet{Cucciati2012}, who cover the largest redshift range with their study. There is a good overall agreement between the confidence region of the model (grey area) and the data points. This shows how efficient our approach is in spite of involving very few scaling laws. We do find that our model is systematically higher than the \citet{Cucciati2012} points at the faint-end between z=0.4 and z=0.8, but this offset is always within the confidence region of the model. This could be caused by a problem of incompleteness in the faintest magnitude bins of the literature studies. Our model is also 2-$\sigma$ higher than \citet{Oesch2010} measurements at L$_{\rm UV}>10^{10}$\,L$_\odot$ in the range $1.7<\rm z<2.5$, but agrees with the other authors in this regime.\\

We switched off the modification of the IRX-M relation at low redshift to quantify its impact. The yellow dashed lines represent the prediction of the model with a non-evolving relation. Between z=0.6 and z=1, there is only a mild impact, since the offset from the constant IRX-M relation at higher redshift is small. At lower redshift, the model without low-z evolution systematically underestimates the bright-end; the faint-end is not affected by changes to the IRX-M relation. Our model thus favors a scenario with a minor evolution of the IRX-M relation at low redshift, implying slightly more transparent galaxies at fixed stellar mass in the local Universe.\\

We also studied the effect of the scatter on the IRX-M relation. The red dotted lines in Fig.\,\ref{fig:uvlf} show the model predictions in absence of scatter. There are almost no changes below the knee of the UV luminosity function, but the model without scatter underpredicts the number of objects ($\sim$0.5\,dex) above the knee of the luminosity function by an order of magnitude. A scatter is thus mandatory to reproduce the bright-end of the UV luminosity function. Indeed, because of the existence of a main-sequence, only intermediate- and high-mass ($\gtrsim \rm M^\star$) galaxies can produce a sufficient SFR and thus intrinsic UV luminosity. However, because of the IRX-M relation, the more massive the object is, the higher its attenuation will be. The scatter generates a population of intermediate mass galaxies with a lower attenuation and thus a bright observed UV luminosity. To test this hypothesis, we selected $\rm L_{UV}>10^{11}\,\rm L_\odot$ galaxies at $1.7< \rm z < 2.5$ in the catalogue based on our best model (with a scatter of 0.4\,dex) and compute their median mass and IRX. We found $6\times10^{10}\, \rm M_\odot$ and 0.81, respectively. The median IRX at this mass is 1.62. UV bright populations are thus M$^\star$ galaxies, which are negative outliers of the attenuation relation. Finally, we tested various values for the scatter on the attenuation relation and found that the UV luminosity function is well reproduced only for scatters in the 0.2-0.5\,dex interval, in agreement with the observational estimates cited in Sect.\,\ref{subsec:refinements}.\\

\begin{figure}
	\centering
	\includegraphics[width=9cm]{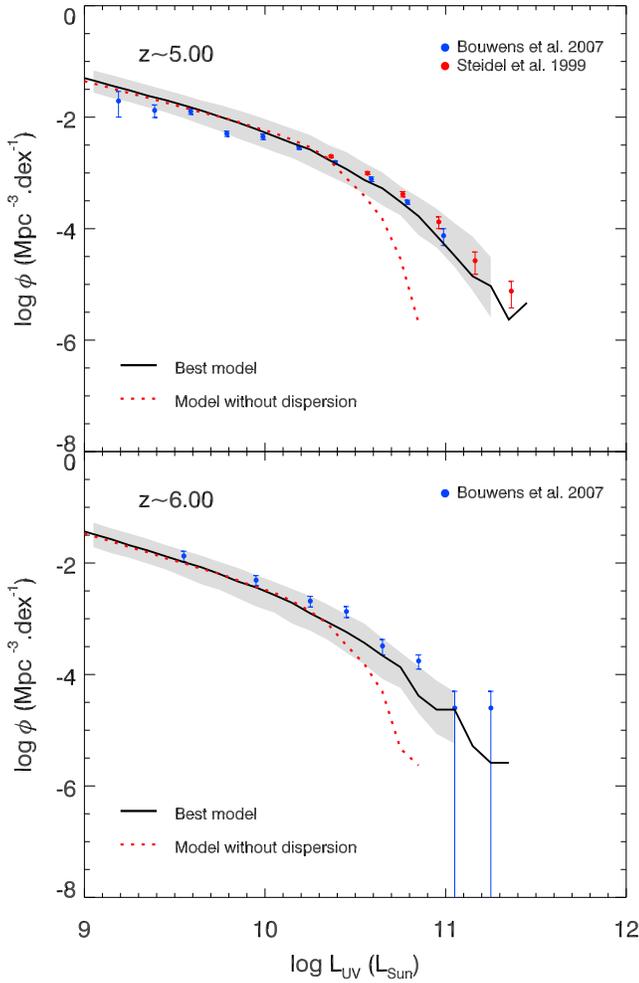}
	\caption{\label{fig:uvlf_lbg} UV luminosity function at z$\sim$5 and z$\sim$6 from LBG-selected samples. Black solid lines represent the UV luminosity function derived from our best model. Red dotted lines represent the same model, but without scatter on the attenuation relation. Colored points are measurements from various observational studies \citep{Bouwens2007, Steidel1999}.}
\end{figure}


\subsection{Comparison with UV luminosity function of high-redshift LBGs}

We extended our model to $z>4$ as a first order test of whether the simple ingredients we assumed at lower redshift are sufficient for reproducing high-redshift measurements as well. Unfortunately, we currently have no access to IR data at these redshifts, but UV luminosity functions were measured using the Lyman-break-selection technique \citep{Steidel1996}. However, our parametric mass function based on \citet{Ilbert2013} was calibrated only up to z=4. We thus used the fit of a compilation of mass function measurements coming from K-band-selected and LBG-selected populations performed by Sargent et al. (in prep.). Basic features of this compilation are a successive steepening of the faint-end slope, a quickly declining characteristic density, and a decreasing M$^\star$ with redshift. We assume no evolution of the sSFR-M and IRX-M relation beyond a redshift of 4 for the sake of simplicity.\\ 

Figure\,\ref{fig:uvlf_lbg} shows the comparison between our model (black solid line) and the data of \citet{Bouwens2007} and \citet{Steidel1999} at z$\sim$5 and 6. All the data points lie in the confidence area of our empirical predictions (grey area). This demonstrates the predictive power of our method and suggests that a non-evolving IRX-M and SFR-M relation at high-z is a fair hypothesis to interpret the current data. In a similar way as at lower redshifts, it is clear that a scatter on the attenuation relation is necessary to reproduce the bright-end of the UV luminosity functions at z$\sim$5 and 6.\\


\begin{figure}
	\centering
	\includegraphics[width=9cm]{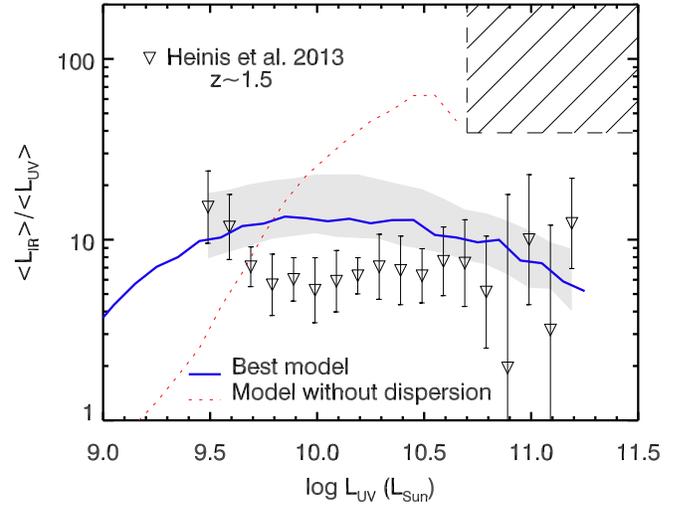}
	\caption{\label{fig:attluv} Mean attenuation as a function of UV luminosity at $z \sim 1.5$. The blue solid line is the prediction of our best model; the grey area is the associated 1$\sigma$ confidence region. The red dotted line stands for the model without scatter in the IRX-M relation. Inverted triangles are from the stacking analysis of \citet{Heinis2013a}. The hatched region indicates the parameter space where no object was found by \citet{Buat2012} in a 0.025\,deg$^2$ field (see discussion Sect.\,\ref{sect:sbobs}).}
\end{figure}

\section{IR properties of UV selected population}

\label{sect:irpropuv}

The IR properties of UV-selected population are not expected to be similar to those of the full population. Our model is a useful tool to understand how this selection biases the obtained results. We now discuss recent observational results obtained on these UV-selected populations, and how they can be interpreted by our model.\\

\subsection{Mean attenuation in UV-selected galaxies}

\citet{Heinis2013a} studied the mean IRX of UV-selected populations, measured with a stacking analysis as a function of $\rm L_{UV}$ at z$\sim$1.5. The results are compatible with a plateau at $10^{\rm IRX}=6.9\pm1.0$ between $10^{9.5}$ and $10^{11}$\,L$_\odot$ (see Fig.\,\ref{fig:attluv}). The flat trend come from the mix of low and high mass galaxies in each UV bin, but is not trivial to understand, because of the rising IRX-M relation \citep{Heinis2013b}. However, our best model also predicts a flat trend in this $\rm L_{UV}$ range (blue solid line on Fig.\,\ref{fig:attluv} ). This result is essentially caused by the scatter on the attenuation relation, since we find a rising trend in absence of dispersion (red dotted line on Fig. \,\ref{fig:attluv}). This is another indication for the importance of the scatter in order to be able to model simultaneously the statistical properties UV and IR populations. At fixed UV luminosity, there is thus a mix between galaxies with higher SFR (mass) and attenuation and others with lower SFR (mass) and attenuation. This picture is consistent with the distribution in the IRX-M diagram of populations selected by UV luminosity recovered from the simulation of \citep{Heinis2013b}. Nevertheless, there is a 2$\sigma$ systematic discrepancy around $10^{10}$\,L$_\odot$. We tried to modify the scatter and the IRX-M relation to reproduce this feature without success. This could be caused by cosmic variance, but also a clue of an higher complexity of real galaxy populations. Finally, our model predicts a rising relation with $\rm L_{UV}$ at $\rm L_{UV}<10^{9.5}$\,L$_\odot$, in contrast with \citet{Heinis2013a} who have found clues of a decreasing trend for their three faintest points. Measurements at lower UV luminosity will thus be a challenging test for our model.\\

\subsection{Absence of strongly-attenuated, UV-bright galaxies}

\label{sect:sbobs} 

The results of \citet{Heinis2013a} were obtained in a stacking analysis and do not provide any information about the dispersion around the IRX-L$_{\rm UV}$ relation. \citet{Buat2012} have studied the distribution of IR- and UV-detected population between z=0.95 and z=2.2 in the same diagram. This type of analysis could be biased by the need of a double detection. However, we note that they found no objects with $\log \,( \rm L_{\rm UV}) \gtrsim  10.7 \, \rm L_{\odot}$ and IRX $\gtrsim 1.6$ (hatched region in fig.\ \ref{fig:attluv}) in a 0.028\,deg$^2$ field, where all galaxies should be detected in both UV and IR. Whereas, on average, SB galaxies represent only $3\%$ of the entire population, they contribute $55 \%$ in this particular area of the IRX-L$_{\rm UV}$ diagram. We will thus use this constraint on the $\rm IRX-\rm L_{\rm UV}$ diagram to test the impact of our model prescription for attenuation in SBs (see Sect.\,\ref{subsec:refinements}). When applying the same attenuation for MS and SB galaxies, we count 1.72 objects ($\rm P[\rm N=0] = 0.17$) in the empty region of the $\rm IRX-\rm L_{\rm UV}$ plane. If we assume that $\rm sSFR_{SB}$ is completely reprocessed into IR we count 0.99 objects ($\rm P[N=0]=0.37$) in the same region. Therefore, observations slightly favour our refined recipe for attenuation in SB. Nevertheless, studies on larger fields are necessary to conclusively test the validity of our assumptions in this regime.\\

\begin{figure*}
	\centering
	\includegraphics[width=17cm]{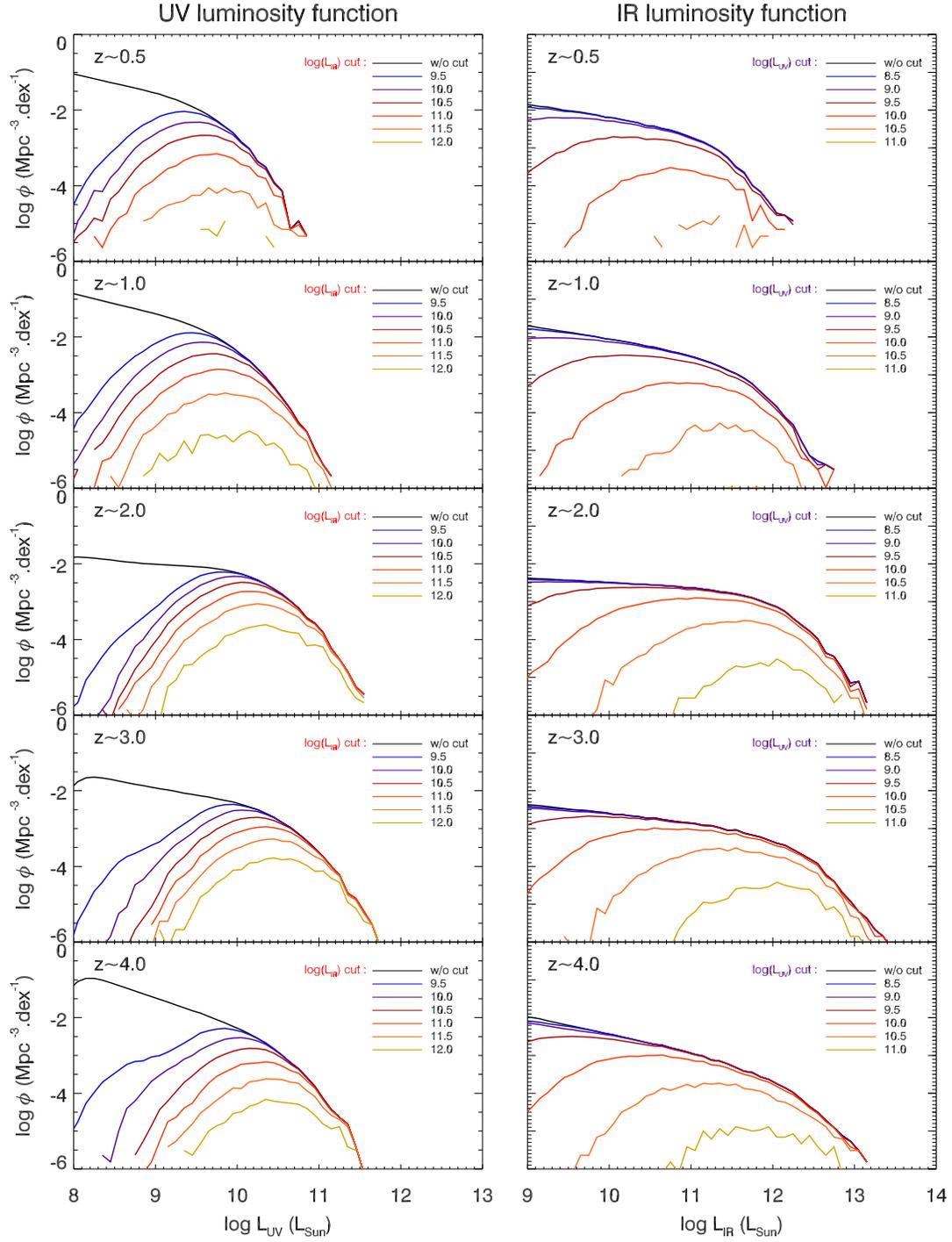}
	\caption{\label{fig:cut} \textbf{Left panels}: Contribution of galaxies above various IR luminosity cuts to the UV luminosity functions at various redshifts.  \textbf{Right panels}: Contribution of galaxies above various UV luminosity cuts to the IR luminosity functions at various redshifts.}
\end{figure*}

\begin{figure*}
\centering
\includegraphics[width=17cm]{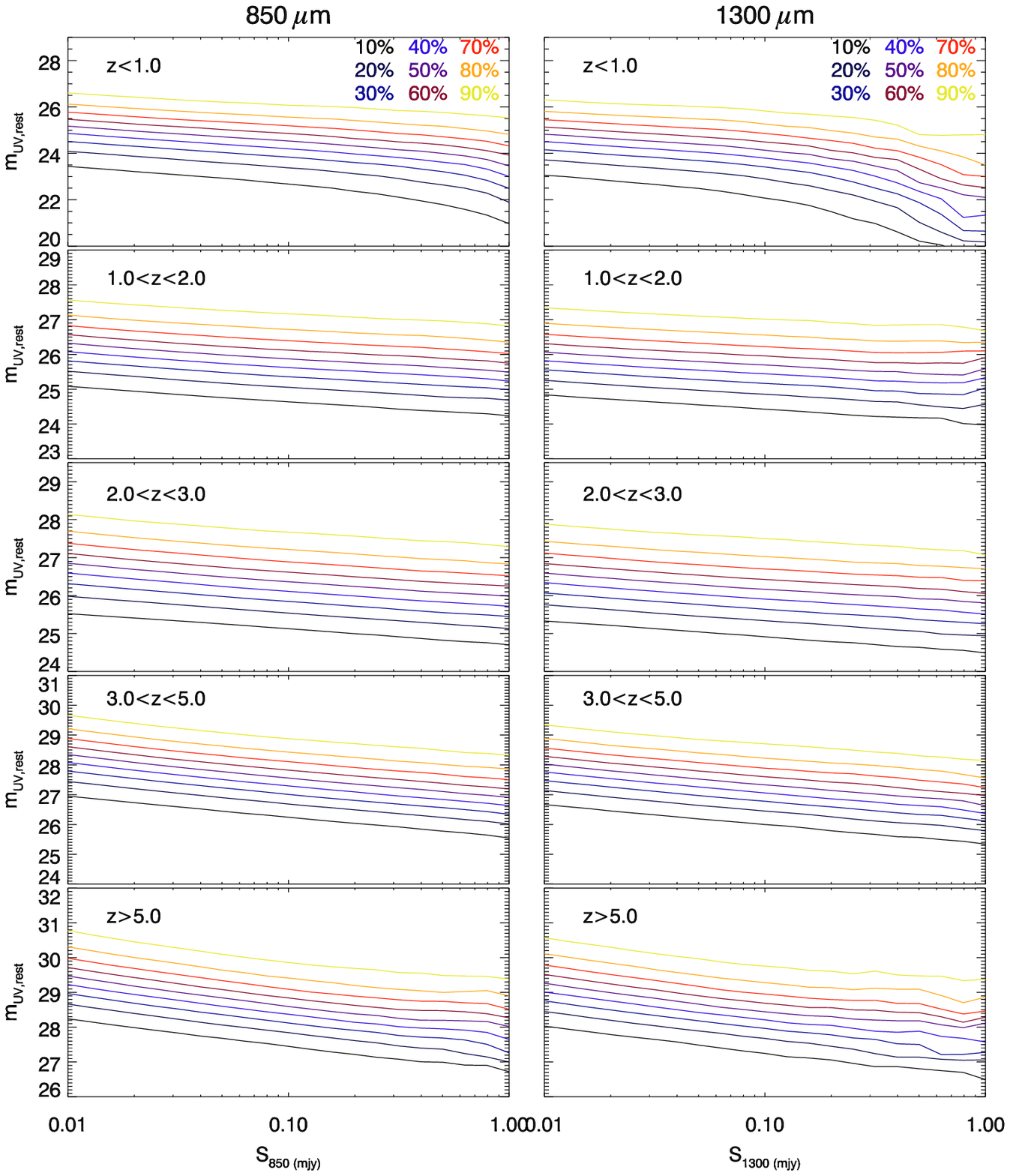}
\caption{\label{fig:fluxalma} AB apparent magnitude at (1+z) $\times$ 150\,nm at which a given fraction of galaxies (colour coded) are detected as a function of their (sub-)millimeter flux. Left panels correspond to fluxes at 850\,$\mu$m and right panels to fluxes at 1300 \,$\mu$m. Various redshift slices are presented, from the lowest on the top panels, to the highest on the bottom panels.}
\end{figure*}

\section{Predictions}

\label{sect:pred} 

All predictions presented in this section are based on our 'full' model, i.e. the realisation with a scatter on IRX-M relation, a special treatment of the starbursts, and an evolution of the IRX-M relation below z=1.

\subsection{Contribution of IR-selected populations to the UV luminosity function}

\label{sect:IRtoUV}

We used our model to predict how much IR galaxies above various luminosity cuts contribute to the UV luminosity function (see Fig.\,\ref{fig:cut} left panel). At all redshifts, we note that the bright-end of the UV luminosity function is predominantly populated by IR-bright galaxies. This is expected, because UV-bright galaxies are essentially M$^*$ galaxies with a median IRX of 0.81 (see Sect.\,\ref{sect:lfuv}). These objects thus mainly emit in the IR. The sub-L$^\star$ regime is populated by IR-faint sources, and ultra deep IR observations are necessary to recover a significant fraction of these objects in the IR. This part of the UV luminosity function is dominated by low-mass galaxies, which have correspondingly low attenuations and low SFRs, and so, low IR luminosities. We also note in the left hand side of Fig.\,\ref{fig:cut} that the UV luminosity distribution of IR-selected sources peaks around the knee of the UV luminosity function for a large range of IR-luminosity cuts ($10^{10}-10^{13}$\,L$_\odot$). The mean UV luminosity of IR-selected population thus slowly evolves as a function of L$_{\rm IR}$ (1\,dex in UV for 2.5\,dex in IR). Indeed, the brighter L$_{\rm IR}$ is, the more massive and attenuated the galaxy will be. Consequently, a strong increase of L$_{\rm IR}$ involves only a weak one of L$_{\rm UV}$, because the highest SFR is partially compensated by an higher attenuation.\\

\subsection{Contribution of UV-selected populations to the IR luminosity function}

We then studied the contribution of UV-selected galaxies to the IR luminosity function (see Fig.\,\ref{fig:cut} right panel). We found that the brightest IR galaxies ($>10\, \rm L_{IR}^\star$) are not the brightest UV galaxies, contrary to what was found for the opposite case in Sect.\,\ref{sect:IRtoUV}. SB galaxies dominates this regime and are very attenuated (see Sect.\,\ref{subsec:refinements}). Consequently, they tend to be missed by shallow UV surveys. In contrast, the sub-L$^\star$ galaxies are well recovered by UV surveys, because they are in general less attenuated. We also note that the L$_{\rm IR}$ distribution of galaxies above a given L$_{\rm UV}$ is much broader than in the opposite case. IR-selected populations have thus more homogeneous L$_{\rm UV}$ than L$_{\rm IR}$ of UV-selected populations. Overall, UV selections are better suited to study low-mass, low-SFR populations, because of their low attenuation. But, IR selections are better suited to study massive, strongly-star-forming galaxies, which tend to be missed by UV surveys because of their strong attenuation. Our model thus confirms the complementarity of UV and IR data, and confirms the commonly-accepted fact that UV is better to probe low SFR and IR is better to probe the high SFR. \\

\subsection{Will the sources of millimeter deep surveys be detected in UV?}

In the next years, ALMA, NIKA on PdBI, and CCAT will perform deep IR continuum surveys around a wavelength of 1\,mm. Detecting the UV rest-frame counterpart of such sources will be useful to constrain the dust attenuation of these distant objects, and better understand the stellar, gas, and dust content of such objects. Our model enables us to estimate the typical depth of rest-frame UV surveys requested to match the sensitivity of future millimeter surveys. Figure\,\ref{fig:fluxalma} shows the fraction of UV-detected sources as a function of the depth of the survey in the millimeter domain and in UV-rest-frame. To simplify the reading of this plot, we defined m$_{\rm UV}$ as the apparent magnitude of a source at 150$\times (1+z)$\,nm. This corresponds to FUV, U, R, I, and J band for sources at z=0, z=1, z=2.65, z=4.35, and z=7.15, respectively. The fluxes in the millimeter band at 850\,$\mu$m and 1.3\,mm are computed using the IR SED templates of \citet{Magdis2012}. This library provides two distinct templates for main-sequence galaxies, and starbursts that depend only on redshift (but not on, e.g., luminosity once the redshift is fixed). Following \citet{Bethermin2012c}, we assume a scatter of 0.2\,dex on the parameter $\langle \rm U \rangle$, which represents the mean intensity of the radiation field and which correlates with dust temperature. Using this SED library, galaxy number counts (including counts split in different redshift bins) from the mid-IR to millimeter wavelengths are correctly reproduced with the 2SFM formalism \citep{Bethermin2012c}.\\

The typical depth of surveys performed at the confusion limit with the next generation of millimeter cameras on thirty-meter class telescopes (NIKA2 at IRAM, LMT, CCAT) will be $\sim$0.5\,mJy \citep{Bethermin2011}. These telescopes will be able to cover fields of 0.1-10\,deg$^{2}$ at this sensitivity. An optical depth of $m_{\rm UV,rest} \sim 26.5$ (COSMOS, \citealt{Capak2007}\footnote{The depth of the GALEX imaging in COSMOS, i.e. of the observer-frame UV-coverage of COSMOS is shallower than quoted here. We use $m_{\rm UV,rest}$ to denote the apparent magnitude at the wavelength which samples the rest-frame UV-emission of galaxies at different redshifts. The bulk of the sources we consider here are at z$>$1 and thus have rest-frame UV emission that falls into observed optical bands, for which the sensitivity is around 26.5 magnitudes.}) will thus provide the UV luminosity to $>80\%$ up to z$\sim$2.5. At z$>$5, a depth of 29\,magnitudes will be necessary. This depth is nowadays reached in field of only few arcmin$^2$. The new generation of millimeter interferometers (ALMA and NOEMA) will be able to probe fluxes one order of magnitude fainter in fields of few arcmin$^2$ or less. In this case, a typical magnitude of 29, which is already reached in HUDF \citep{Coe2006}, will allow to detect the $>80\%$ of UV counterparts at z$<$5. At larger redshift, a depth of $\sim30$ will be necessary. This should be possible with the James Webb Space Telescope (JWST).\\

\section{Conclusion}

\label{sect:conc}

We modeled the evolution of the UV and IR emissions of galaxies across cosmic times based on the 2SFM formalism, built from the empirically-measured evolution of the main-sequence of star-forming galaxies, and the observed IRX-M relation assuming a 0.4\,dex scatter around it. This modeling work allows to better understand the connection between the IR-selected and UV-selected populations. Our main findings are:
\begin{itemize}
\item Our model is able to consistently predict the co-evolution of the IR luminosity function up to z$\sim$3 and the UV luminosity function up to z$\sim$6.
\item We showed that scatter on the IRX-M relation has no effects on the IR luminosity function. This is not the case for the UV, where a scatter of $\sim$0.4\,dex is necessary to reproduce the UV luminosity function. This is caused by the fact that the UV bright galaxies are not the most massive or the most star-forming, but M$^*$ galaxies, which are negative outliers of the attenuation relation.
\item We recover naturally a flat mean IRX-$\rm L_{UV}$ relation measured by stacking in \textit{Herschel} data of UV-selected sources as a consequence of the scatter on the IRX-M relation, the shape of the mass function, and the SFR-M relation.
\item We performed predictions from our model and showed that IR-selected populations have similar $\rm L_{UV}$  around the knee of the UV luminosity function, while UV-selected populations can have very different IR properties. Overall, the IR-selection is very efficient for selecting massive and/or strongly-star-forming galaxies, but UV is much better suited for studying the low-mass, weakly star-forming galaxies.
\end{itemize}
This simple model will be useful to understand the selection bias of various studies of star-formation in distant galaxies, and especially the IR-selected sample from the next generation of deep interferometer surveys or large single-dish surveys.\\

\section{acknowledgements}
We thank E. Le Floc'h, D. Elbaz, and Olivier Ilbert for useful/helpful comments, and S. Arnouts, O. Cucciati, N.P. Hathi, P.A. Oesch, and T.K. Wyder for providing data. EB thanks C. Schreiber for help with IDL programming. We aknowledge the anonymous referee for their constructive and helpful comments. MB, MS, and ED acknowledge support from ERC-StG UPGAL 240039 and ANR-08-JCJC-0008.\\

\bibliographystyle{mn2e}

\end{document}